# Anomalously Strong 2D Band Intensity in Twisted Bilayer Graphene: Raman Evidence for Doubly Degenerate Dirac Band

Yanan Wang,[1,†] Zhihua Su,[1,†] Wei Wu,[1,2] Shu Nie,[3] Xinghua Lu,[4] Haiyan Wang,[5] Kevin McCarty,[3] Shin-shem Pei,[1,2] Francisco Robles-Hernandez,[6] Viktor G. Hadjiev,[7] Jiming Bao[1,*]

[1]Department of Electrical and Computer Engineering
University of Houston, Houston, TX 77204

[2]Center for Advanced Materials
University of Houston, Houston, TX 77204

[3]Sandia National Laboratories, Livermore, CA 94550

[4]Institute of Physics, Chinese Academy of Sciences, Beijing 100190, China

[5]Department of Electrical and Computer Engineering
Texas A&M University, College Station, Texas 77843

[6]College of Engineering Technology
University of Houston, Houston, TX 77204

[7]Texas Center for Superconductivity and Department of Mechanical Engineering, University of Houston, Houston, TX 77204

# Abstract

We report the observation of anomalously strong 2D band in twisted bilayer graphene (tBLG) with large rotation angles under 638-nm and 532-nm visible laser excitation. The 2D band of tBLG can reach four times as opposed to two times as strong as that of single layer graphene. The same tBLG samples also exhibit rotation dependent G-line resonances and folded phonons under 364-nm UV laser excitation. We attribute this 2D band Raman enhancement to the constructive quantum interference between two double-resonance Raman pathways which are enabled by nearly degenerate Dirac band in tBLG Moiré superlattices.

Graphene has been the focus of attention for nearly a decade since the first successful fabrication of single layer film by mechanical exfoliation.[1] This is because besides many other interesting properties, graphene also provides a clean and versatile platform that allows us to test various theories with experiments. As a result, basic properties of single-layer and AB-stacking multi-layer graphene have been well understood.[2] Among many graphene characterization techniques, Raman scattering has proven to be a powerful non-invasive method that can probe graphene band structure and subsequently distinguish single layer from multi-layer graphene.[2,3] The strong Raman signal is made possible by graphene unique cone-like band structure and Raman resonance in which the exciting laser can always match the graphene interband transition.[3] As a two dimensional atomically thin film, graphene also provides a new degree of freedom that is not possessed by other nanostructures: relative rotation between graphene layers.[4] Rotationally twisted bilayer graphene (tBLG) has recently attracted intensive theoretical and experimental investigations.[4-21] Rotation dependent properties such as van Hove singularity,[12] G-line resonances[5,8,10] and folded phonons have been revealed.[14,20,22,23] However, unlike AB-stacking graphene, twisted bilayer graphene remains as a challenge to researchers. Chemical synthesis of bilayer graphene with a controlled rotation angle has not been reported, and many properties of tBLG, especially with non-commensurate lattices have not been well understood.

In this letter, we report the observation of anomalously strong 2D band in tBLG with large rotation angles. In contrast to the expected two-fold enhancement,[5] its intensity is about four times as strong as that in monolayer graphene under visible laser excitation. We attribute such enhanced 2D band to quantum interference between two Raman pathways enabled by interlayer coupling and degenerate Dirac band in tBLG. The observation of enhanced 2D Raman as well as G-line resonances and folded phonons indicates that twisted bilayer graphene can be effectively described by Moiré superlattice.[7,14,20-22] A clear picture of band structure is essential to the understanding of other properties of twisted bilayer graphene.

Twisted bilayer graphene was grown on Cu foils by CVD at ambient pressure in a quartz tube furnace.[15,22,24,25] Fig. 1 shows scanning electron microscopy (SEM) images of four bilayer graphene samples and their Raman spectra under 638-nm laser excitation. These samples have increasing rotation angles from 9 to 16 degrees, and the expected rotation dependent Raman spectrum can be clearly seen.[5,8,22] G-line resonances can only be observed in samples with smaller 9° and 11° rotation angles. The folded phonon can also be seen in Fig. 1b near 1500 cm$^{-1}$. In addition, the 2D intensity and line width have also experienced significant changes: the integrated intensity increases but the line width decreases as the rotation angle increases, as summarized in Fig. 2a.

These rotation dependent Raman spectra of G line and 2D band agree with previously reported observations and calculation.[5,8,10] However, there are two noticeable differences compared with previous work, especially with the calculation[5]. First, the 2D intensity becomes more than twice that of single layer graphene. Second, the line width drops below ~40 cm$^{-1}$ of single layer 2D band. Such observations are general in our bilayer graphene with rotation angle larger than 20 degrees. Fig. 2b-c show typical Raman spectra of two bilayer graphene with rotation angles around 25 degrees. As can be seen, the intensity of G line is doubled, but the 2D band intensity of bilayer graphene becomes about four times stronger, and the line width of 2D becomes ~5 cm$^{-1}$ narrower. As a comparison, for "artificial" bilayer graphene made from simple mechanical stacking by transferring monolayer graphene twice, the 2D band intensity is approximately twice that observed in single-layer graphene, and the line width remains same, as shown in Fig. 2d.

The observation of a narrower but stronger 2D band in twisted bilayer graphene is not limited to 638-nm excitation laser. Fig. 3a shows Raman spectra under 532-nm laser from the same sample as in Fig. 2c. A strong 2D band is also observed. But when excited by a 364-nm UV laser, a G-line resonance is observed again, and the 2D band becomes weaker and broader, as shown in Fig. 3b.[22] It is important to point out that the peak with frequency very close to the D-line of single layer graphene is not the D-line of bilayer graphene, instead, it is the folded longitudinal optical (LO) phonon of tBLG superlattice.[22] In addition to a good match between its frequency with the rotation angle,[22,23] the following observation can rule out this peak as the D band of bilayer graphene: it is red-shifted but the 2D band shows a blue-shift when compared to the corresponding D and 2D bands of single layer graphene. If this peak is the D-line of tBLG, it is expected to shift in the same direction as 2D band. As a matter of fact, both D and 2D bands in previous Raman spectra in Fig. 1 and 2 exhibit the same blue-shifts in bilayer graphene.

Figure 4 shows four more examples of enhanced 2D band in tBLG with large rotation angles near 25° under the 638-nm laser excitation. Again we can see that the intensity of G-line doubles but the intensity of 2D band almost quadruples. It should be noted that all these samples show G-line resonances and folded LO phonons under the 364-nm UV excitation.[22] Because all the samples were obtained from the same batch and Raman measurements were performed under the same condition, the only variable among the samples is the rotation angle, therefore this rotation dependent 2D band Raman behavior must be a reflection of the intrinsic property of tBLG. Other major factors such as doping and strain have been shown to affect graphene Raman spectrum, but they can be ruled out in our case. For example, the enhanced 2D band cannot be a result of rotation dependent strain in tBLG.[26] The strain is negligible because both G line width and line shape remain the same for both single layer and bilayer graphene. Previous study also shows that the strain has no strong effect on the intensity and line width of 2D band.[26] Recent calculation indicates that twisted bilayer graphene does not show a preferred rotation angle,[21] so a significant strain is not expected from the relative rotation between graphene layers. An enhanced 2D band was reported in suspended single layer graphene, and it was attributed to decreased doping level in the suspended region[27]. This doping effect can also be excluded based on the following reasons. First, as discussed before that there is no noticeable change in G line width. Second, the 2D band is blue-shifted instead of red-shifted.[27]

We believe this enhanced 2D band intensity in tBLG is a strong indication of constructive quantum interference between two Raman pathways.[28-31] Such enhancement as well as G-line resonances and folded phonons cannot be observed in the artificially stacked bilayer graphene (sBLG) where interlayer coupling can be ignored. This is an essential difference between tBLG and sBLG, and this difference holds the key to the unique Raman features of tBLG. Fig. 5 shows schematics of band structures of tBLG and sBLG along with representative 2D band double resonance Raman pathways.[3,6,10,19] Here we have assumed that both types of bilayer graphene films form two dimensional superlattices defined by Moiré pattern[7,13,14,20,32]. For comparison, we only show Raman pathways that involve interband transition between two inner Dirac bands. Let's take a look at the difference between Raman pathways. In tBLG shown in Fig. 5a, we have two Raman pathways A→B→D and A→C→D as in AB-stacking bilayer graphene. But in sBLG shown in Fig. 5b, the Raman pathway A→C→D is not allowed. This is because the interband transition A->C is forbidden due to the lack of interlayer coupling in sBLG. In other words, the inner and outer Dirac bands represent bands of individual graphene layers. Cross-interlayer excitation is negligible if there is no interlayer interaction.

Let's examine the consequence of different Raman pathways on the Raman spectra of G-line and 2D band. First, we note that the absorption of an incident photon through interband transition $P_{int}$ is the same for both types of bilayer graphene. Compared with single layer graphene, the paths for interband transition are doubled for both bilayer graphene. As a result, the G-line intensity doubles as the laser excitation spot moves from single layer to bilayer region. But for 2D band, we can see that the Raman pathway after A splits into two in tBLG while there is only one Raman pathway in sBLG. In the latter case, the Raman amplitude is proportional to $P_{int}*P_{ABD}$. But in the case of tBLG, the transition amplitude is $P_{int}* (P_{ABD}+ P_{ACD})/sqrt(2)$. Note that the square root 2 sqrt(2) is due to the even splitting of incident amplitude at the point A. Assuming $P_{ABD}$ and $P_{ACD}$ have the same amplitude and phase due to nearly degenerate Dirac band,[3,6,9,11,19,33] the transition amplitude can be written as $P_{int}*P_{ABD}*sqrt(2)$. Because Raman intensity is the modular square of transition amplitude, the 2D Raman intensity in tBLG becomes twice the Raman in sBLG, or in other words, four times that of single-layer graphene.

It should be noted that the quantum interference is enabled by the unique degenerate Dirac band structure of tBLG, which has been calculated for twisted bilayer graphene with commensurate lattice by many groups.[6,8-11] In these cases, the degeneracy of Dirac band is almost guaranteed because bilayer graphene becomes a two dimensional superlattice with larger unit cells. The band structure of such superlattice can be constructed by zone folding of monolayer graphene band structure into reduced Brillouin zone. The band below the M point, or G-line resonance, is degenerate Dirac band. For twisted bilayer graphene with an incommensurate lattice, there is no simple general theoretical description of band structure. However, the observation of rotation dependent folded phonons indicates that bilayer graphene can be approximated as a Moiré superlattice[7,13,14,20,32].

It is important to note that there is a finite gap (~20 meV for 13.2 degrees tBLG) between two Dirac bands[6]. This band splitting is much smaller than ~300 meV in AB-stacking graphene. Besides G-line resonances and folded LO phonons, the low frequency breathing mode of tBLG is another indication of interlayer coupling[34]. It is the finite interlayer coupling and finite gap between Dirac bands that makes it possible to observe such quantum interference between two Raman pathways. For a band splitting of ~20 meV, it is estimated that Raman peak spacing between two paths shown in Fig. 5 is on the order of 2-5 cm$^{-1}$, which is much smaller than measured ~40 cm$^{-1}$ line width of 2D-band[3,6]. In other words, the two Raman paths are indistinguishable, which is the necessary condition for quantum interference. For a larger gap as in AB-stacking bilayer graphene, peaks from such two Raman pathways can be well resolved, and two Raman paths are distinguishable, so there is no such quantum interference[3]. When there is no interlayer coupling as in sBLG, two graphene layers are independent from each other, Raman intensity of any lines should then be a simple sum of Raman intensity from the individual layers, or twice the Raman intensity of a single layer as shown in Fig. 2d.

The stronger but narrower 2D band in tBLG was observed before, but it is not fully understood yet[5,10]. The narrower line width is not due to the improved lattice quality in tBLG. It is well known that a higher lattice quality will reduce the D line intensity, but in our case, the D-line is also enhanced as 2D band. We believe this enhanced D line is also related to the quantum interference because there are degenerate Raman pathways for D-line. The exact cause for the narrower 2D band width requires further investigation.

In conclusion, we have observed enhanced 2D band Raman in tBLG under visible laser excitations, and established its correlation with G-line resonances and folded phonons under UV excitation. The enhancement is due to Raman quantum interference enabled by double resonance and degenerate Dirac band of tBLG Moiré superlattices. Quantum interference has been observed in first-order Raman scattering in semiconductors and carbon nanotubes[28-31,35,36], but it has not been reported in second-order double resonance Raman scattering. Our findings help to reveal more novel properties and applications of twisted bilayer graphene.

**Acknowledgements**

The work at Sandia National Laboratories was supported by the Office of Basic Energy Sciences, Division of Materials Sciences and Engineering of the US DOE under contract No. DE-AC04-94AL85000. SSP, JMB, and WW acknowledge support from the Delta Electronics Foundation and UH CAM. JMB acknowledges support from the National Science Foundation (Career Award ECCS-1240510 monitored by Anupama Kaul, DMR-0907336 monitored by Charles Ying) and the Robert A Welch Foundation (E-1728). VGH work was supported by the State of Texas through the Texas Center for Superconductivity at the University of Houston.

†These authors contributed equally to this work.

*jbao@uh.edu

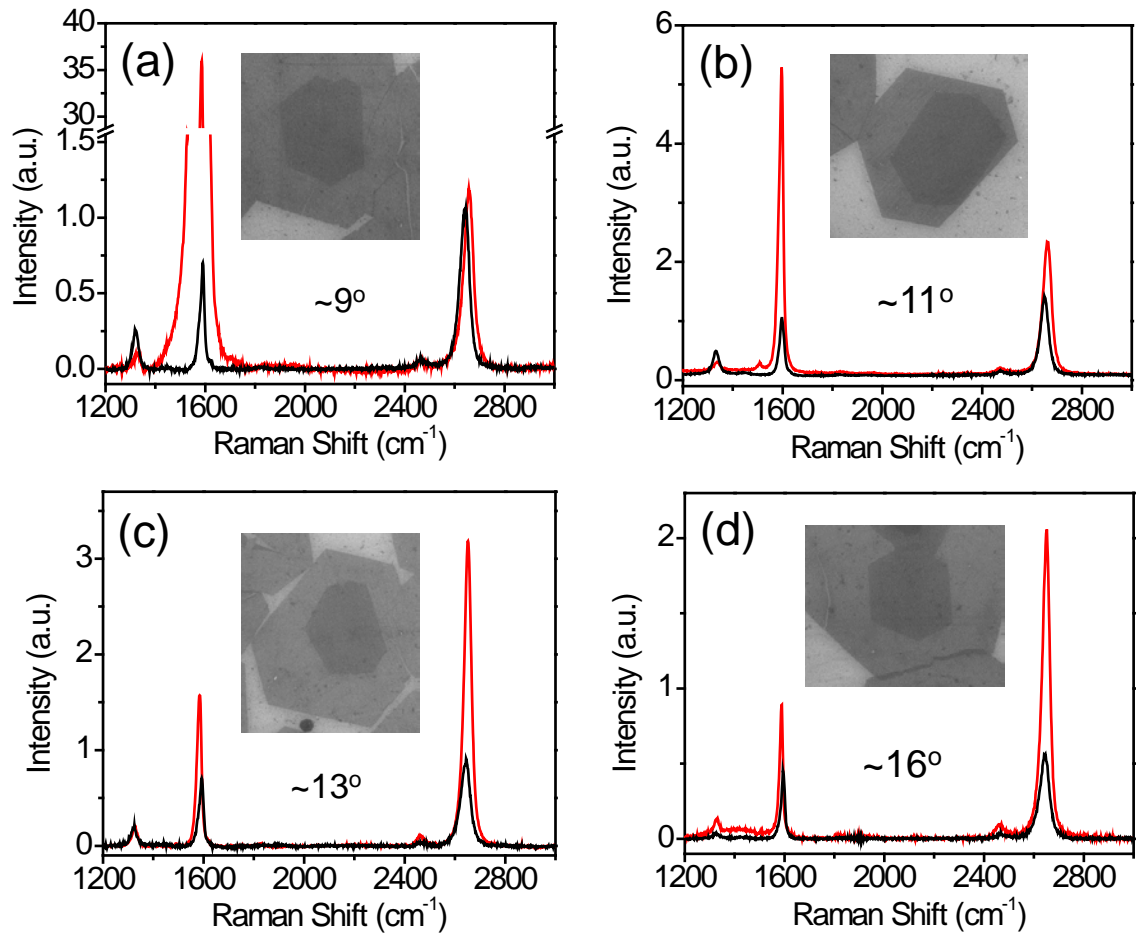

FIG. 1. (Color online) Raman spectra of single layer graphene (in black) and twisted bilayer graphene (in red) with rotation angles of 9, 11, 13 and 16 degrees. Insets are scanning electron microscopy (SEM) images. The excitation laser wavelength is 638 nm.

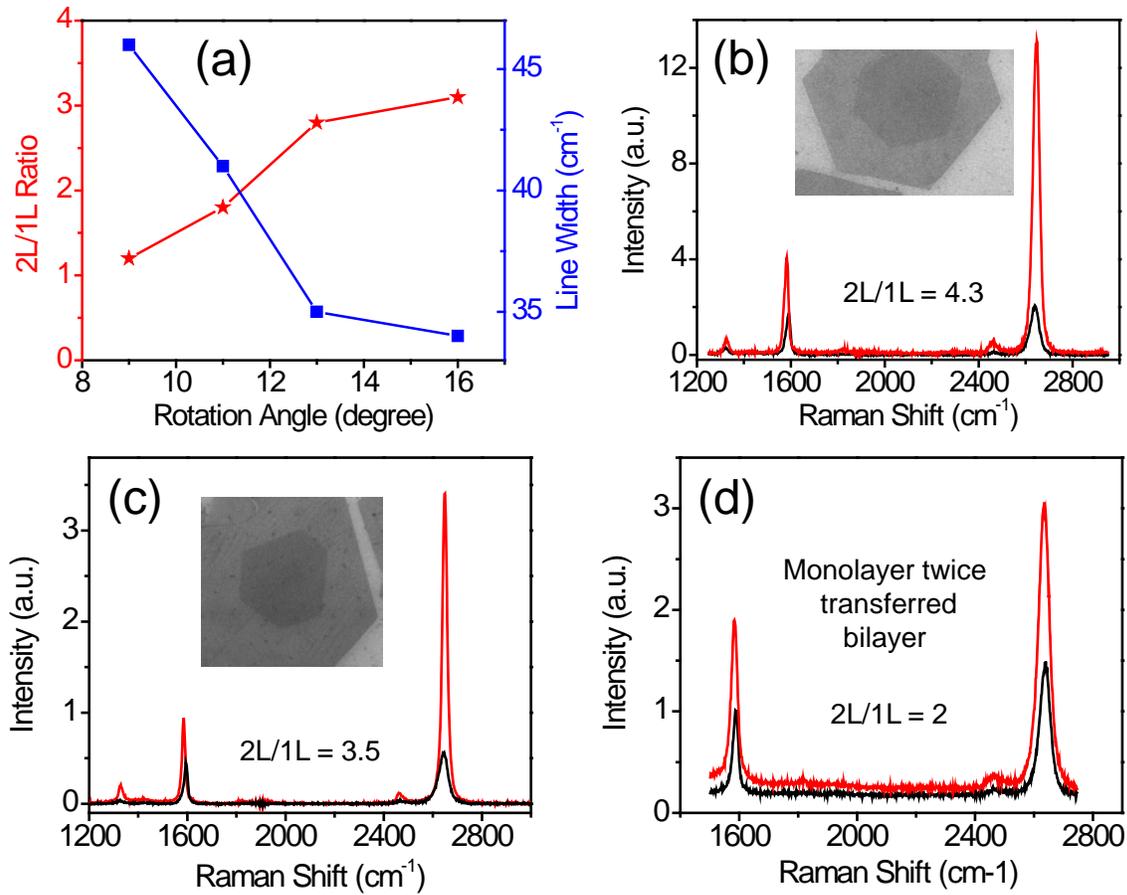

FIG. 2. (Color online) (a) Evolutions of 2D band line width and normalized 2D band intensity of bilayer versus single layer graphene as a function of rotation angle. (b-d) Raman spectra of single layer (in black) and bilayer (in red) graphene. Insets in (b) and (c) are SEM images of tBLG with large rotation angles near ~25°. The wavelength of laser is 638 nm. Bilayer graphene in (d) was fabricated by transferring one monolayer graphene onto another monolayer graphene. "2L/1L" is 2D band integrated intensity ratio of bilayer to single layer graphene.

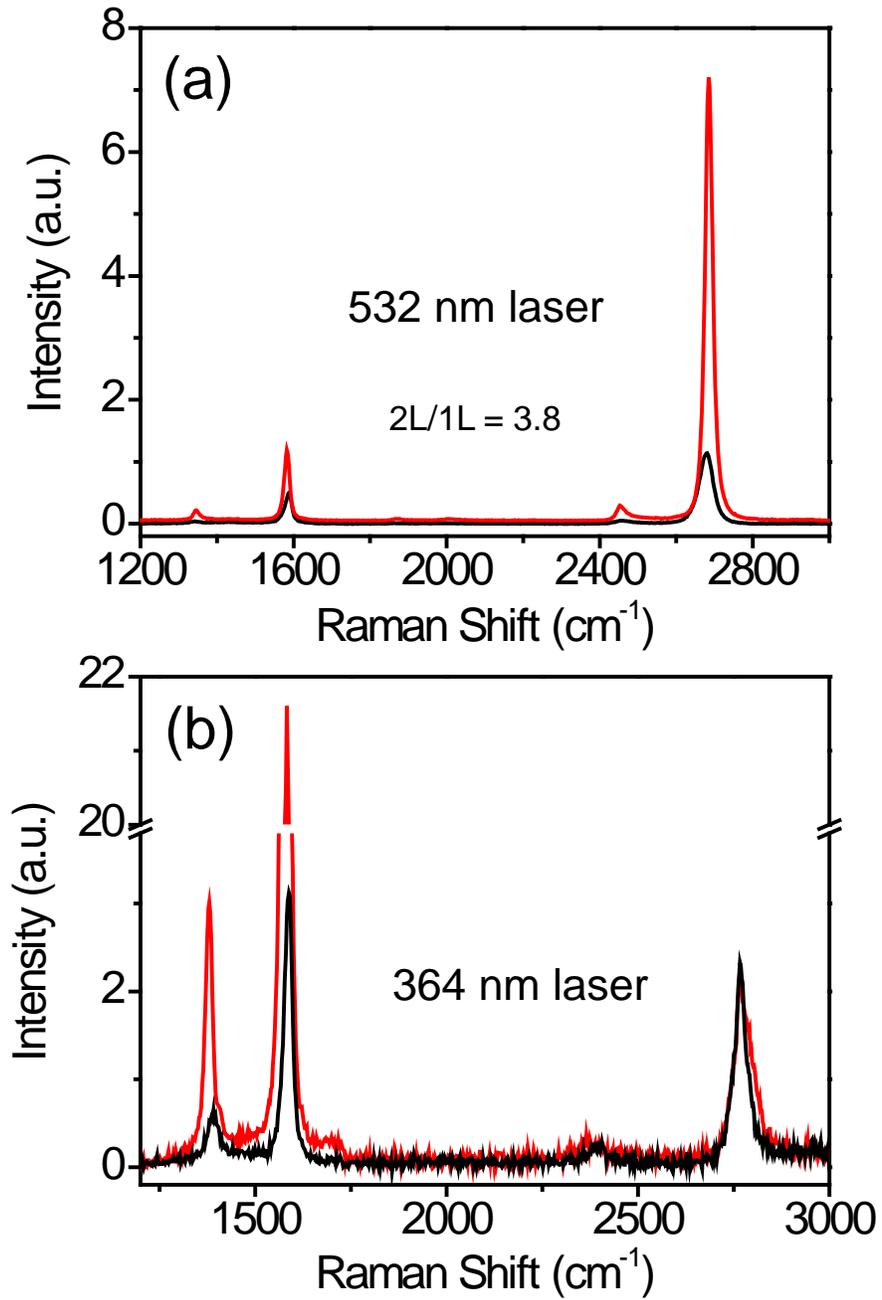

FIG. 3. (Color online) Raman spectra of the sample shown in Fig. 2c under 532-nm (a) and 364-nm (b) laser excitations. Single layer and bilayer Raman spectra are plotted in black and red respectively.

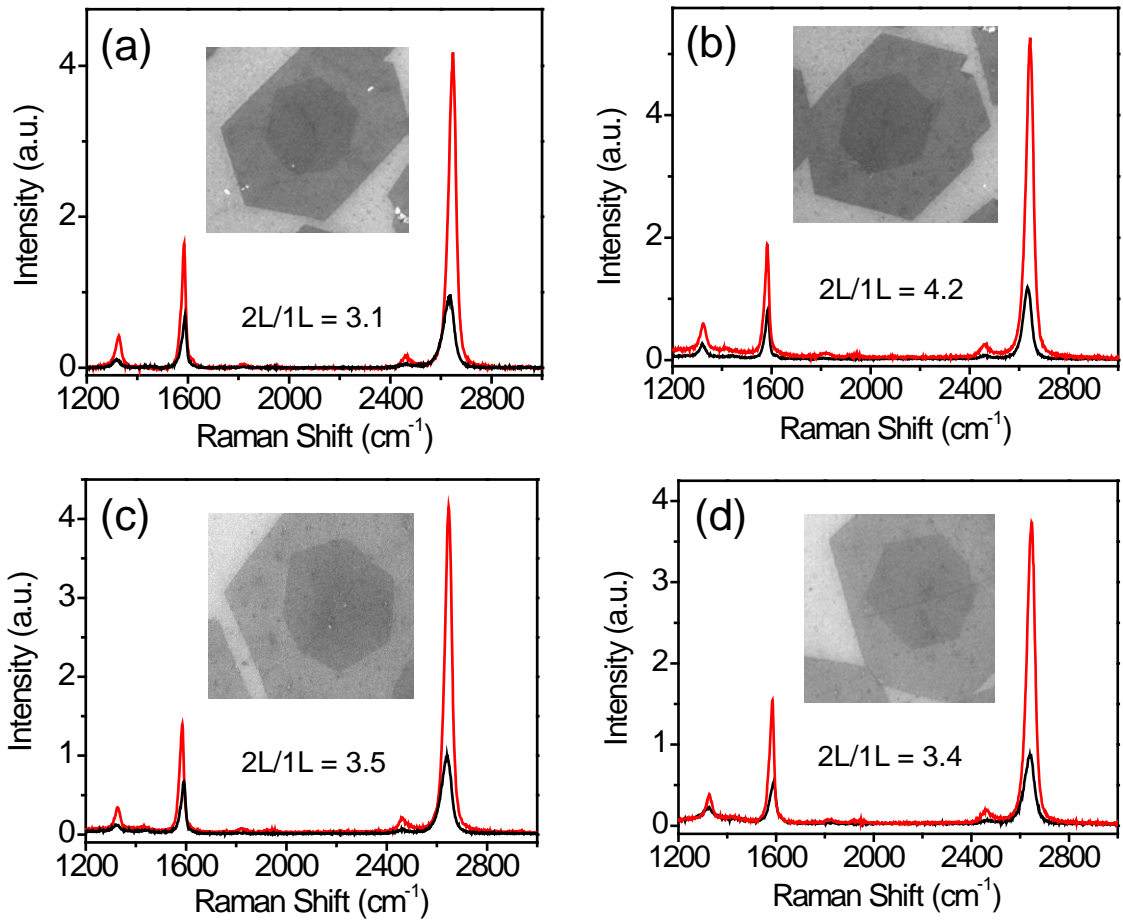

FIG. 4. (Color online) Same as in Fig. 1 except that graphene samples have large rotation angles (>20 degrees).

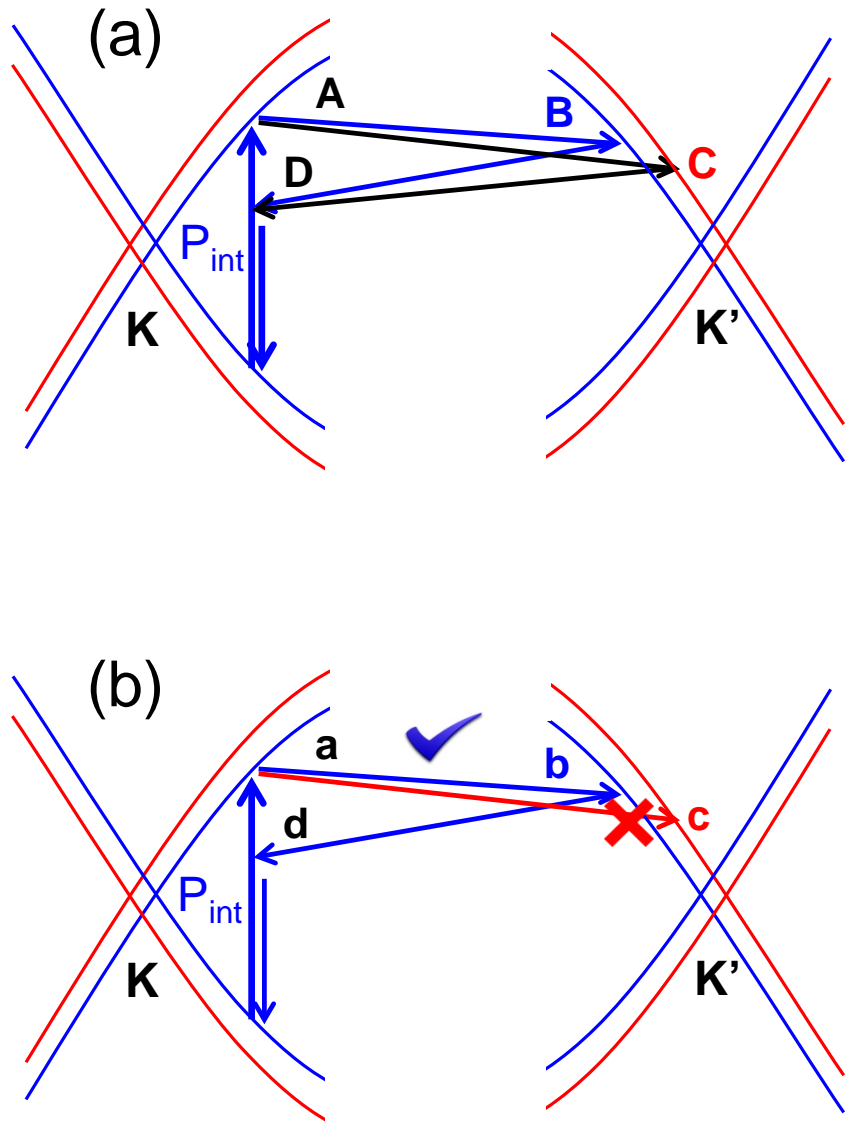

FIG. 5. (Color online) Schematics of representative 2D-band double-resonance Raman pathways in (a) tBLG and (b) bilayer graphene with weak or no interlayer coupling. Only Raman contributions from inner loop and transitions involving electrons are shown.